\documentclass {elsart}
\usepackage {epsfig}

\begin {document}

\begin {frontmatter}

% use the thanksref command within \title, \author or \address for
% footnotes;
% use the corauthref command within \author for corresponding author
% footnotes;

\title {Dimensionality reduction in translational noninvariant wave guides}

\author {Khee-Kyun Voo\thanksref {Acknowledgement}} 
% \ead [url] {home page}
\corauth [cor] {Tel.: +886-2-77380145. fax: +886-2-77387411}
\ead {kkvoo@mail.oit.edu.tw}
\thanks [Acknowledgement] {Supported by the National Science Council of
Taiwan under Grant NSC96-2112-M-161-001.}

%\author [label1] {Khee-Kyun Voo}
% \address [label1] {}
% \address [label2] {}

\address {Department of Communication Engineering, Oriental Institute of
Technology, Taipei county 220, Taiwan}

\date {\today}

\begin {abstract}

A scheme to reduce translational noninvariant quasi-one-dimensional
wave guides into singly or multiply connected one-dimensional (1D) lines
is proposed. It is meant to simplify the analysis of wave guides, with
the low-energy properties of the guides preserved. Guides comprising
uniform-cross-sectional sections and
discontinuities such as bends and branching junctions are considered. The 
uniform sections are treated as 1D lines, and the discontinuities are
described by equations sets connecting the wave functions on the lines.
The procedures to derive the equations and to solve reduced systems are
illustrated by examples, and the scheme is found to apply when the
discontinuities are distant and the energy is low. When the scheme
applies, it may substantially simplify the analysis of a wave guide, and
hence the scheme may find uses in the study of related problems, such as
quantum wire networks.

\end {abstract}

\begin {keyword}  Wave guide, discontinuity, quantum wire, one-dimensional,

\PACS  43.20.Mv,73.23.Ad,73.63.Nm,84.40.Az

\end {keyword}

\end {frontmatter}

\section {Introduction}
\label {intro}

When a wave propagates with a low energy in a quasi-one-dimensional (Q1D)
wave guide with an uniform cross section, the guide is effectively
one-dimensional (1D) since only the first transverse mode plays a role.
However, when a guide comprises discontinuities such as bends and
branching junctions, higher transverse modes come into play at the
vicinity of the discontinuities. Therefore, in principle, a full-mode or
full-wave analysis is required, which means a substantial increase in the
amount of calculation and the physics is often obscured. Knowing that the
guide is essentially 1D in the uniform sections, such a full-wave
analysis for the entire system is actually unnecessary. It is the purpose
of this paper to present a scheme to eliminate those redundancies and
simplify the calculations, yet have those low-energy properties faithfully
preserved.

Historically, there has been a number of schemes related to this purpose.
The earliest one was due to Kuhn and dates back to 1949 \cite {Kuh49}. 
Then it was made more well-known by Griffith in 1953 \cite {Gri53}, and
henceforth it has been called the Griffith boundary condition \cite
{RS53,ES89,KS99,Xia92,MHZ93,DJ94,Mos97,RC98,BDJ04,BGC04,FMB05,AWS05,WV05,KFB08}.
The scheme contains a set of equations relating the wave functions and
their first derivatives on the lines connected to a junction \cite
{Kuh49,Gri53,RS53,ES89,KS99,Xia92,MHZ93,DJ94,Mos97,RC98,BDJ04,BGC04,FMB05,AWS05,WV05,KFB08}.
The equations are simple equations that satisfy the unitarity condition at
the junction. The scheme has been intuitively stated and has no
undetermined parameters. However, recently it has been pointed out by the
author $et~al.$ \cite {VCT06} that it is not clear what kind of realistic
Q1D guides the scheme describes. Later than Kuhn, there was another scheme
by Shapiro in 1983 \cite {Sha83}. The scheme starts by an unitary matrix
relating the amplitudes of the inward and outward waves \cite
{Sha83,BIA84,GIA84}. Though this scheme contains free parameters, it also
has not been mentioned how the parameters are related to the guides in 
realistic spaces.

To the present, approaches to the related problems are belong to either
one of the mentioned two categories \cite
{Kuh49,Gri53,RS53,ES89,KS99,Xia92,MHZ93,DJ94,Mos97,RC98,BDJ04,BGC04,FMB05,AWS05,WV05,KFB08,VCT06,Sha83,BIA84,GIA84}.
and a common feature of most of the schemes \cite
{Kuh49,Gri53,RS53,ES89,KS99,Xia92,MHZ93,DJ94,Mos97,RC98,BDJ04,BGC04,FMB05,AWS05,WV05,KFB08,Sha83,BIA84,GIA84}
is that the relation between the reduced systems and the original systems is
not addressed, and this hinders the application to the study of realistic
systems.

In 2006, the author {\it et al.} proposed a scheme \cite {VCT06},
which has been the first attempt to relate reduced and unreduced systems.
The scheme resembles that by Kuhn \cite {Kuh49}, but has an extra
phenomenological term with a tuning parameter. The parameter is
to be fixed by a comparison between results from the reduced and
unreduced systems. The scheme was shown to be a substantial improvement,
and many low-energy transport properties were shown to be captured.
But still, it is pointed out later in this paper that this simple
phenomenology can be inadequate and a more general scheme is needed.

Section \ref {form} illustrates the derivation of the connecting equations
to be used in reduced systems, for two typical component structures ---
the L-bend and the T-junction. Then in Sec.~ \ref {examples}, these two
structures are assembled into more elaborated structures, and the wave
propagation in the structures are studied in and compared between, the
reduced and unreduced systems. In Sec.~\ref {conc}, a few concluding
remarks are given.

\section {Formulation}
\label {form}

The scheme to be proposed can be summarized as in the following. For a
particular discontinuity, the scattering matrix ($S$-matrix) is evaluated
using a full-wave treatment, and then the matrix is truncated leaving only
those elements relating the first transverse modes in the branch guides.
The truncated $S$-matrix is then used to connect the 1D wave functions in
reduced systems.

The two-dimensional (2D) time-independent Schr$\ddot {\rm o}$dinger
equation (TISE) $-[ \hbar^2 / (2m_0)] [ \partial_x ^2 \Psi + \partial_y ^2
\Psi ] + V(x,y) \Psi = E \Psi$, with $V(x,y)=0$ and hard wall boundaries
is considered. For a wave with an energy $E$ in
a guide having a width $W$, the wave function can be written as a
sum of direct products of transverse modes and longitudinal waves. 
Labeling the guide by $\eta$ and defining a coordinate system $(x_\eta, 
y_\eta)$ in the guide, where $x_\eta$ is in the longitudinal direction
and $y_\eta$ is in the transverse direction ($0 < y_\eta < W$), the wave
function can be written as
\begin {eqnarray}
\displaystyle
\Psi_\eta (E; x_\eta,y_\eta) &=& \sum _{m=1}^{N_\eta} 
\sqrt {2 \over W} ~ {\rm sin} \left( { {m \pi y_\eta} \over W} \right)
\nonumber \\
&& \times 
\left(  A^{(m)}_\eta e^{ik^{(m)} x_\eta} + B^{(m)}_\eta e^{-ik^{(m)}
x_\eta} \right),
\label {psi_2d}
\end {eqnarray}
where $k^{(m)} \equiv \sqrt {2m_0 E/\hbar^2 - (m \pi / W )^2 }$ is the
longitudinal wave number for the $m$-th transverse mode, which can be
propagating or evanescent, and $N_\eta$ is a large enough integer. In a 1D
space, the TISE becomes $-[\hbar^2 / (2m_0)] \partial_x ^2 \psi + V(x)
\psi = E \psi$. For a line labeled by $\eta$, and with a coordinate
$x_\eta$ defined on it, the wave function for $V(x)=0$ is
\begin {eqnarray}
\psi_\eta (E; x_\eta, y_\eta) = 
A_\eta e^{ik x_\eta} + B_\eta e^{-ik x_\eta},
\label {psi_1d}
\end {eqnarray}
where $k$ is the longitudinal wave number given by $k = \sqrt {2m_0 E} /
\hbar$. When results from 2D guides and 1D lines are compared,
$A^{(1)}_\eta$ ($B^{(1)}_\eta$) is compared with $A_\eta$ ($B_\eta$), and 
$k^{(1)}$ is compared with $k$.

\begin {figure}

\includegraphics{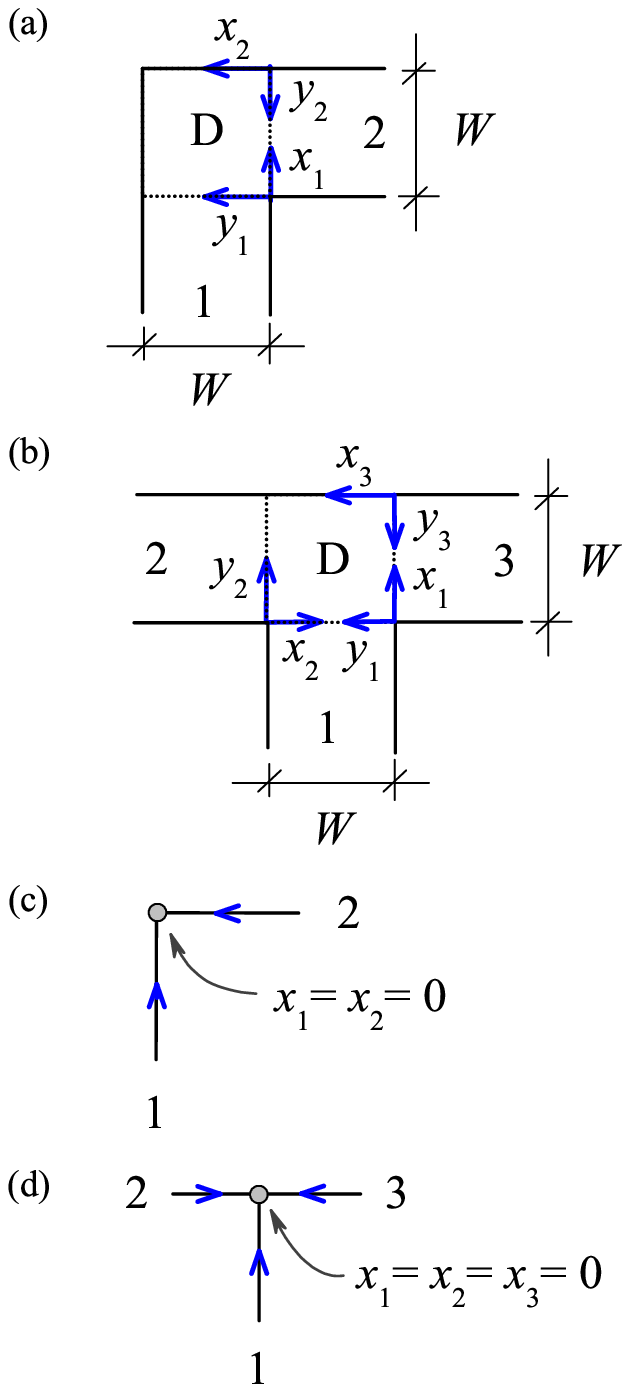}
\vspace {15cm}

\caption {
(Color online)
The two discontinuities in 2D wave guides considered in this paper,
(a) the L-bend and (b) the T-junction. The widths of the guides are
denoted by $W$. The branch guides are labeled by 1, 2, and 3, in which
coordinates are defined as shown. The translational noninvariant
regions are labeled by D. In the reduced systems, (c) a L-bend is
represented by a ``L,'' and (d) a T-junction is represented by a
``T.'' The arrows on the branch lines indicate the positive directions of
the coordinates defined on the lines, and the origins of the coordinates
are defined to be at the branching nodes.
}

\label {discontinuities}
\end {figure}

Two discontinuities in 2D wave guides are considered, the L-bend and
T-junction as shown in Fig.~\ref {discontinuities}(a) and \ref
{discontinuities}(b) respectively. The discontinuities are divided into
regions which are labeled by 1, 2, 3, and D in our discussion. Coordinates
are also defined in the branch guides, and wave functions in the branch
guides are in the form given by Eq.~(\ref {psi_2d}). The $S$-matrices are
to be presented in terms of these coordinates. The symbols used to denote
the L-bend and T-junction in reduced systems are shown in Fig.~\ref
{discontinuities}(c) and \ref {discontinuities}(d) respectively.

We only sketch the evaluation of the $S$-matrices here, since the
techniques are well-established and are detailed in the literatures \cite
{Dat95,FG97}. The wave function in region D [see Figs.~\ref
{discontinuities}(a) and \ref {discontinuities}(b)] is also expanded in
terms of undetermined amplitudes, and they are connected to wave functions
in other regions by the conditions of continuities of wave functions and
normal derivatives of wave functions at the boundaries between the
regions. Otherwise, one may also solve the TISE in a discretized space,
where the TISE is a set of finite-difference (FD) equations.

For the T-junction shown in Fig.~\ref {discontinuities}(b), we can get an
equation for the amplitudes of the waves in the branch guides (regions 1,
2, and 3) such as
\begin {eqnarray}
\left[
\begin {array} {c}
B^{(1)}_1 \\ B^{(1)}_2 \\ B^{(1)}_3 \\ ...
\end {array}
\right]
=
\left[
\begin {array} {cccc}
s_{11} ^{\rm T} & s_{12} ^{\rm T} & s_{13} ^{\rm T} & ... \\
s_{21} ^{\rm T} & s_{22} ^{\rm T} & s_{23} ^{\rm T} & ... \\
s_{31} ^{\rm T} & s_{32} ^{\rm T} & s_{33} ^{\rm T} & ... \\
... & ... & ... & ... \\
\end {array}
\right]
\left[
\begin {array} {c}
A^{(1)}_1 \\ A^{(1)}_2 \\ A^{(1)}_3 \\ ...
\end {array}
\right],
\label {st_2d}
\end {eqnarray}
where $A^{(m)}_\eta$ ($B^{(m)}_\eta$) is an amplitude for an inward
(outward) wave in Eq.~(\ref {psi_2d}). Truncating the matrices in
Eq.~(\ref{st_2d}) and retaining only terms related to the first transverse 
modes in the branch guides, we get
\begin {eqnarray}
\left[
\begin {array} {c}
B_1 \\ B_2 \\ B_3
\end {array}
\right]
=
S _{\rm T} \left[
\begin {array} {c}
A_1 \\ A_2 \\ A_3
\end {array}
\right],
\label {Teqs}
\end {eqnarray}
where $A_\eta$ and $B_\eta$, $\eta=1,2,$ and 3 are amplitudes in
Eq.~(\ref {psi_1d}), and
\begin {eqnarray}
S _{\rm T} =
\left[
\begin {array} {ccc}
s_{11} ^{\rm T} & s_{12} ^{\rm T} & s_{13} ^{\rm T} \\
s_{21} ^{\rm T} & s_{22} ^{\rm T} & s_{23} ^{\rm T} \\
s_{31} ^{\rm T} & s_{32} ^{\rm T} & s_{33} ^{\rm T} 
\end {array}
\right],
\label {st_1d}
\end {eqnarray}
where the coordinates are chosen as toward the junction, and $x_1 = x_2 =
x_3 = 0$ at the junction as shown in Fig.~\ref {discontinuities}(d). 
For the scheme to work, it is necessary that the energy is such that only
the first transverse mode is propagating, and the exponential tails of the
evanescent waves of higher modes emanated from the discontinuities are
shorter than the distances between the discontinuities.

We can also write Eq.~(\ref {Teqs}) in terms of $\psi_\eta$. Using
$A_\eta = [ \psi_\eta + d\psi_\eta / d(ikx_\eta) ] / 2 $ and $B_\eta
= [ \psi_\eta - d \psi_\eta /d(ikx_\eta) ] / 2 $ at $x_\eta = 0$,
the connecting equation can be rewritten as
\begin {eqnarray}
\displaystyle { {1 + S _{\rm T} } \over {ik } }
\left[
\begin {array} {c}
\displaystyle {{d\psi_1} \over {dx_1}} \\
\displaystyle {{d\psi_2} \over {dx_2}} \\
\displaystyle {{d\psi_3} \over {dx_3}}
\end {array}
\right]
=
\left( 1 - S _{\rm T} \right)
\left[
\begin {array} {c}
\psi_1 \\ \psi_2 \\ \psi_3
\end {array}
\right],
\label {st_psi}
\end {eqnarray}
where the wave functions and their derivatives are evaluated at the
discontinuity, and the directions of the coordinates are defined to be
toward the discontinuity. Note that in Eq.~(\ref {st_psi}), it is not
necessary that the origins of the coordinates be located at the
discontinuity.

Likewise, the connecting equation for the L-bend shown in Fig.~\ref
{discontinuities}(a) can be written as
\begin {eqnarray}
\displaystyle { { 1 + S _{\rm L} } \over {ik} }
\left[
\begin {array} {c}
\displaystyle {{d \psi_1} \over {d x_1}} \\
\displaystyle {{d \psi_2} \over {d x_2}}
\end {array}
\right]
=
\left( 1 - S _{\rm L} \right)
\left[
\begin {array} {c}
\psi_1 \\ \psi_2
\end {array}
\right],
\label {sl_psi}
\end {eqnarray}
where the directions of the coordinates are defined to be toward the
discontinuity as shown in Fig.~\ref {discontinuities}(c).

Note that, the $S$-matrices are symmetric (${S_{\rm L}}^{\rm
t} = S_{\rm L}$ and ${S_{\rm T}}^{\rm t} = S_{\rm T}$ ) and
unitary (${S_{\rm L}} ^\dagger S_{\rm L} = 1$ and ${S_{\rm T}}
^\dagger S_{\rm T} = 1$). The unitarity implies $\sum_\eta (|A_\eta|^2 -
|B_\eta|^2) = 0$, and
since $A_\eta = [ \psi_\eta + d\psi_\eta /d(ik x_\eta) ] / [2 e^{ik
x_\eta} ] $ and $B_\eta = [ \psi_\eta - d\psi_\eta/d(ik x_\eta) ] / [2
e^{-ik x_\eta} ] $, the unitarity can be rephrased as a more
intuitive equality $\sum_\eta \psi_\eta^* d\psi_\eta / d(ik x_\eta) = 0$,
or there is no net inflow of probability current to the discontinuity.

\begin {figure}

\includegraphics{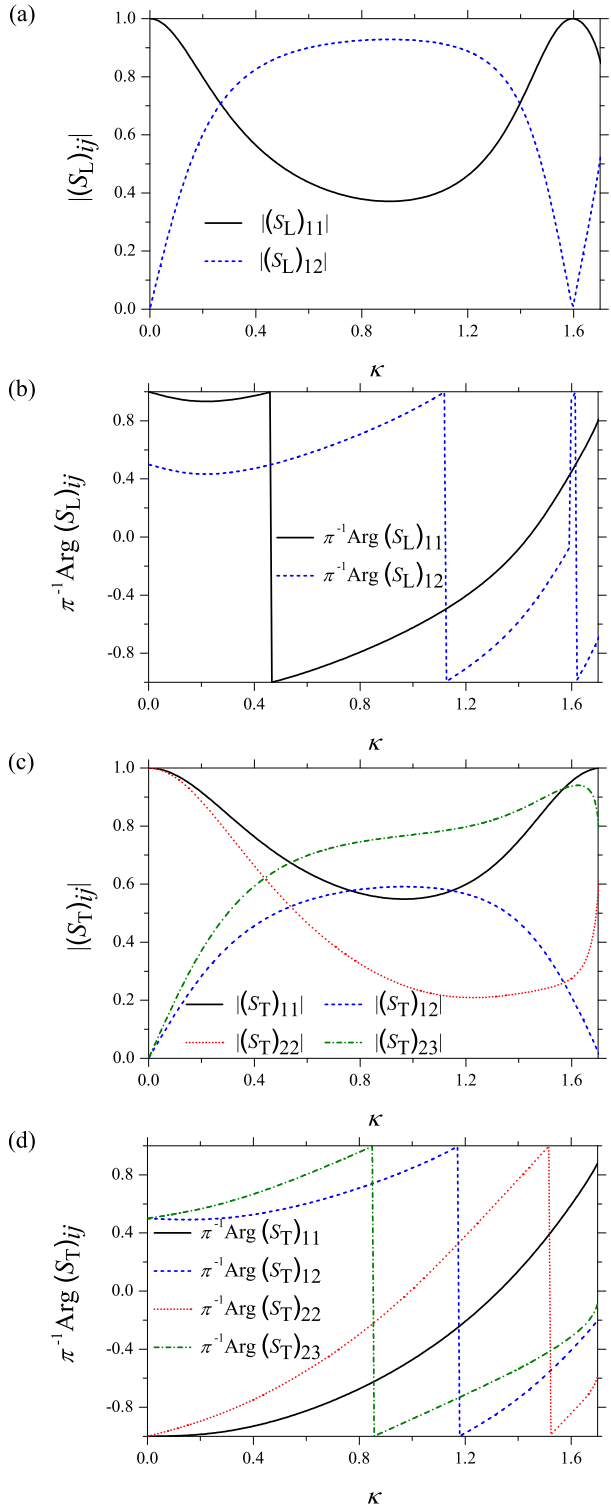}
\vspace {20cm}

\caption {
(Color online)
The (a) magnitudes of the elements of $S_{\rm L}$,
(b) arguments of the elements of $S_{\rm L}$,
(c) magnitudes of the elements of $S_{\rm T}$, and
(d) arguments of the elements of $S_{\rm T}$ are plotted versus a
dimensionless longitudinal wave number $\kappa$, defined by $\kappa
\equiv k^{(1)}W/\pi$.
}

\label {smat}
\end {figure}

The numerical results for the magnitudes and arguments of the elements of
$S_{\rm L}$ are plotted in Figs.~\ref {smat}(a) and \ref {smat}(b)
respectively, and the results for $S_{\rm T}$ are plotted in Figs.~\ref
{smat}(c) and \ref {smat}(d) respectively, versus a dimensionless longitudinal wave
number $\kappa$ defined by $\kappa \equiv k^{(1)}W/\pi$. 
For the reference of the readers, the cutoff of the second
transverse mode is at $\kappa = \sqrt 3 \simeq 1.73$. The results are
obtained using discretized spaces with 20 sites across a width of $W$,
and they are found to be in congruence with results from 10 sites within
windows of $\Delta \kappa \simeq 0.02$ on the horizontal axes, and windows
of $\Delta |S| \simeq 0.02$ and $\Delta [\pi^{-1} {\rm Arg} (S)] \simeq
0.005$ on the vertical axes, which implies that the continuous space limit 
has been approached.

Since the L-bend and T-junction appear quite often in practical 
problems, it may be convenient to have their $S$-matrices in analytic
forms. Within a finite range of $\kappa$, it is possible to
approximate a $S$-matrix by analytic functions. For the $S_{\rm L}$ at
$\kappa < 1$, we may approximate the magnitudes and arguments of the
elements by
\begin {eqnarray}
\displaystyle |(S_{\rm L})_{11}| &\simeq&  {1 \over {1 + (2.796 \kappa -
1.498 \kappa^2 )^2}}, \label {sl11mag_approx} \\
{\rm Arg} ~ (S_{\rm L})_{11} &\simeq& ( 1 - 0.586 \kappa + 1.496
\kappa^2 - 0.541 \kappa^3 ) \pi , \label {sl11arg_approx} \\
|(S_{\rm L})_{12}| &=& \sqrt { 1 - | (S_{\rm L}) _{11} |^2 }, \label
{sl12mag} \\
{\rm Arg} ~ (S_{\rm L})_{12} &=&  {\rm Arg} ~ (S_{\rm L})_{11} -
{\pi \over 2} , \\
(S_{\rm L})_{21} &=& (S_{\rm L})_{12}, ~~ {\rm and} \\
(S_{\rm L})_{22} &=& (S_{\rm L})_{11}.
\label {sl_approx}
\end {eqnarray}
Within $\kappa < 1$, the approximation for $|(S_{\rm L})_{11}|$ in
Eq.~(\ref {sl11mag_approx}) and the approximation for ${\rm Arg} ~ (S_{\rm
L})_{11}$ in Eq.~(\ref {sl11arg_approx}) approximate the numerical
results in
Figs.~\ref {smat}(a) and \ref {smat}(b) respectively up to $\Delta |S| <
0.01$ and $\Delta [ \pi^{-1} {\rm Arg} (S) ] < 0.01$ accuracy. The exact
equalities Eqs.~(\ref {sl12mag})-(\ref {sl_approx}) are due to the
symmetry and unitarity of the $S$-matrix and the exchange of leads 1 and
2.

For the $S_{\rm T}$ at $\kappa < 1$, it is found that the numerical
results for the elements can be approximated by
\begin {eqnarray}
\displaystyle |(S_{\rm T})_{11}| &\simeq&  {1 \over {1 + (1.734 \kappa -
0.808 \kappa^2 )^2}}, \label {st11mag_approx} \\
{\rm Arg} ~ ( S_{\rm T})_{11} &\simeq& (1 - 0.051 \kappa + 0.559 \kappa
^2 + 0.018 \kappa ^3 ) \pi , \\
|(S_{\rm T})_{12}| &=& \sqrt { { 1 - |(S_{\rm T})_{11}|^2 } \over 2 }, \\
{\rm Arg} ~ (S_{\rm T})_{12} &\simeq& \left( {1 \over 2} - 0.153 \kappa +
0.585 \kappa ^2  -0.087 \kappa ^3 \right) \pi , \\
(S_{\rm T})_{13} &=& (S_{\rm T})_{12}, \\
|(S_{\rm T})_{22}| &\simeq& {1 \over {1 + (1.780 \kappa +
0.015 \kappa^2 )^2}}, \label {st22mag_approx} \\
{\rm Arg} ~ (S_{\rm T})_{22} &\simeq& (1 + 0.415 \kappa + 0.508 \kappa
^2 + 0.089 \kappa ^3 ) \pi , \\
|(S_{\rm T})_{23}| &=& \sqrt { { { 1 + |(S_{\rm T})_{11}|^2 } \over 2 }
- |(S_{\rm T})_{22}|^2 } , \\
{\rm Arg} ~ (S_{\rm T})_{23} &\simeq& \left( {1 \over 2} + 0.205 \kappa
+ 0.616 \kappa ^2  - 0.202 \kappa ^3 \right) \pi , \\
(S_{\rm T})_{33} &=& (S_{\rm T})_{22}, ~~ {\rm and} \\
{S_{\rm T}}^{\rm t} &=& S_{\rm T}.
\end {eqnarray}
For $\kappa < 1$, the above approximations for $|S|$ and ${\rm Arg} ~ (S)$
approximate the results in Figs.~\ref {smat}(c) and \ref 
{smat}(d) respectively up to $\Delta |S| < 0.01$ and $\Delta [ \pi^{-1}
{\rm Arg} (S) ] < 0.01$ accuracy. The exact equalities are due to the
symmetry and unitarity of the $S$-matrix, and the exchange of 
leads 2 and 3.

\section {Comparison between reduced and unreduced systems} 
\label {examples}

In this section, three wave guides which are composites of the
discussed L-bend and T-junction are analyzed. The scattering amplitudes
from the original 2D structures and the reduced multiply-connected 1D
structures are compared. Cases of far apart and close discontinuities,
different orientations of guides, and straight and smoothly curved
guides are considered.

\begin {figure}

\includegraphics{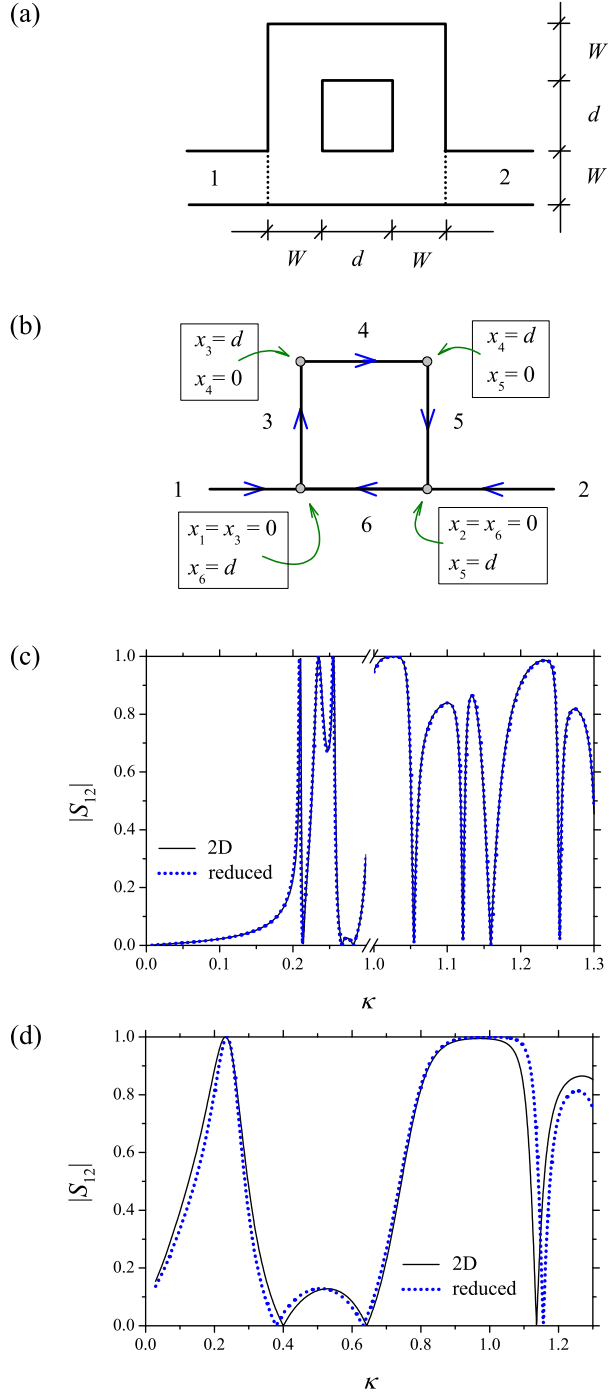}
\vspace {19cm}

\caption {
(Color online)
(a) A considered 2D composite structure. The uniform-cross-sectional
sections have the same width $W$. The distances between the
discontinuities are chosen to be the same and are denoted by $d$.
(b) A reduced version of the above structure. A coordinate $x_\eta$ is
defined on line $\eta$ ($\eta = 1 - 6$) with a positive direction
indicated by an arrow on the line. The coordinates of the discontinuities
on the lines are shown in the boxes.
(c) For $d=4W$, the magnitude of the transmission scattering amplitude
$|S_{12}|$ is plotted versus $\kappa$, for the original 2D system (solid
line) and the reduced system (dotted line). The $\kappa$ here is defined
by $\kappa \equiv k^{(1)}W/\pi$ or $\kappa \equiv kW/\pi$. Seen in the
size of the present graph, the two curves are almost indistinguishable.
(d) For $d=0.1W$, results for $|S_{12}|$ are also shown. The curves
deviate from each other, especially when $\kappa$ gets larger.
}

\label {loop1}
\end {figure}

The first example is a 2D square loop resonator with two leads as depicted
in Fig.~\ref {loop1}(a). The translational invariant sections 
have the same width $W$. For simplicity, the distances between the
discontinuities are chosen to be the same, and are denoted by $d$ as shown.
In Fig.~\ref {loop1}(b), a reduced version for the
structure in Fig.~\ref {loop1}(a) is shown. The magnitude of the transmission 
scattering amplitude $|S_{12}|$ has been plotted versus $\kappa$ for the
unreduced and reduced systems, in Figs.~\ref {loop1}(c) and \ref
{loop1}(d), for $d=4W$ and $d=0.1W$ respectively. The $S_{12}$ for the 2D
structure is the scattering amplitude from the first transverse mode in
lead 2 to the first transverse mode in lead 1, and the dimensionless wave
number $\kappa$ is now defined by $\kappa \equiv k^{(1)}W/\pi$ or $\kappa
\equiv kW/\pi$ depending on the context. The procedure for a full-wave
evaluation of $|S_{12}|$ for the 2D structure is standard \cite
{Dat95,FG97} and it is not to be repeated here, but only the result is 
given. The result here is obtained with a FD TISE, and the number of
sites across a width $W$ is equal to 20.

In the reduced system shown in Fig.~\ref {loop1}(b), a line labeled by
$\eta$ and given a coordinate $x_\eta$ has a wave function in the form given
in Eq.~(\ref {psi_1d}). For the coordinates defined in Fig.~\ref
{loop1}(b), the wave functions on the lines are connected by
\begin {eqnarray}
\displaystyle { { 1 + S _{\rm L} } \over {ik} }
\left[ \begin {array} {c}
\left. ~ \displaystyle {{d\psi_3} \over {dx_3}} \right| _{x_3=d}  \\
\left. - \displaystyle {{d\psi_4} \over {dx_4}} \right| _{x_4=0} 
\end {array} \right]
=
\left( 1 - S _{\rm L} \right)
\left[ \begin {array} {c}
\left. \psi_3 \right| _{x_3=d} \\
\left. \psi_4 \right| _{x_4=0}
\end {array} \right],  
\label {node1} \\
\displaystyle { { 1 + S _{\rm L} } \over {ik} }
\left[ \begin {array} {c}
\left. ~ \displaystyle {{d\psi_4} \over {dx_4}} \right| _{x_4=d}  \\
\left. - \displaystyle {{d\psi_5} \over {dx_5}} \right| _{x_5=0} 
\end {array} \right]
=
\left( 1 - S _{\rm L} \right)
\left[ \begin {array} {c}
\left. \psi_4 \right| _{x_4=d} \\
\left. \psi_5 \right| _{x_5=0}
\end {array} \right],
\label {node2} \\
\displaystyle { { 1 + S _{\rm T} } \over {ik} }
\left[ \begin {array} {c}
\left. - \displaystyle {{d\psi_3} \over {dx_3}} \right| _{x_3=0}  \\
\left. ~ \displaystyle {{d\psi_1} \over {dx_1}} \right| _{x_1=0}  \\
\left. ~ \displaystyle {{d\psi_6} \over {dx_6}} \right| _{x_6=d} 
\end {array} \right]
=
\left( 1 - S _{\rm T} \right)
\left[ \begin {array} {c}
\left. \psi_3 \right| _{x_3=0} \\ \left. \psi_1 \right| _{x_1=0} \\ \left.
\psi_6 \right| _{x_6=d}
\end {array} \right], 
\label {node3} 
\end {eqnarray}
and
\begin {eqnarray}
\displaystyle { { 1 + S _{\rm T} } \over {ik} }
\left[ \begin {array} {c}
\left. ~ \displaystyle {{d\psi_5} \over {dx_5}} \right| _{x_5=d}  \\
\left. ~ \displaystyle {{d\psi_2} \over {dx_2}} \right| _{x_2=0}  \\
\left. - \displaystyle {{d\psi_6} \over {dx_6}} \right| _{x_6=0} 
\end {array} \right]
=
\left( 1 - S _{\rm T} \right)
\left[ \begin {array} {c}
\left. \psi_5 \right| _{x_5=d} \\ \left. \psi_2 \right| _{x_2=0} \\
\left. \psi_6 \right| _{x_6=0}
\end {array} \right]. \label {node4a}
\end {eqnarray}
This contains 10 equations with 10 unknowns, when $A_1$ and $A_2$ are
given. The scattering amplitude $S_{12}$ is obtained as $S_{12}=B_1$ at
$A_1=0$ and $A_2=1$. The magnitude $|S_{12}|$ is plotted in Figs.~\ref
{loop1}(c) and \ref {loop1}(d), for values of $d$ corresponding to the original 
2D structure. In the calculation, $S_{\rm L}$ and 
$S_{\rm T}$ use the numerical values shown in Fig.~\ref {smat}.

In the case of $d=4W$ [see Fig.~\ref {loop1}(c)], results from the
original 2D and the reduced systems are nearly indistinguishable when
seen in the size of the plot, while in the case of $d=0.1W$ [see Fig.~\ref 
{loop1}(d)], the two results have a perceptible difference, especially
when $\kappa$ becomes large. A criterion for the applicability of the
one-mode reduction scheme
is $|k^{(2)} d| \gg 1$, which means that the exponential tails of the
evanescent waves for the higher transverse modes emanating from the
discontinuities, is much shorter than the distances between the
discontinuities. In this regime, the evanescent waves from neighboring
discontinuities do not overlap each other, and the discontinuities
communicate with each other only via the first transverse modes in the
translational invariant sections. Hence, the higher transverse modes in
the sections are redundant, and a single-mode description is adequate.

\begin {figure}

\includegraphics{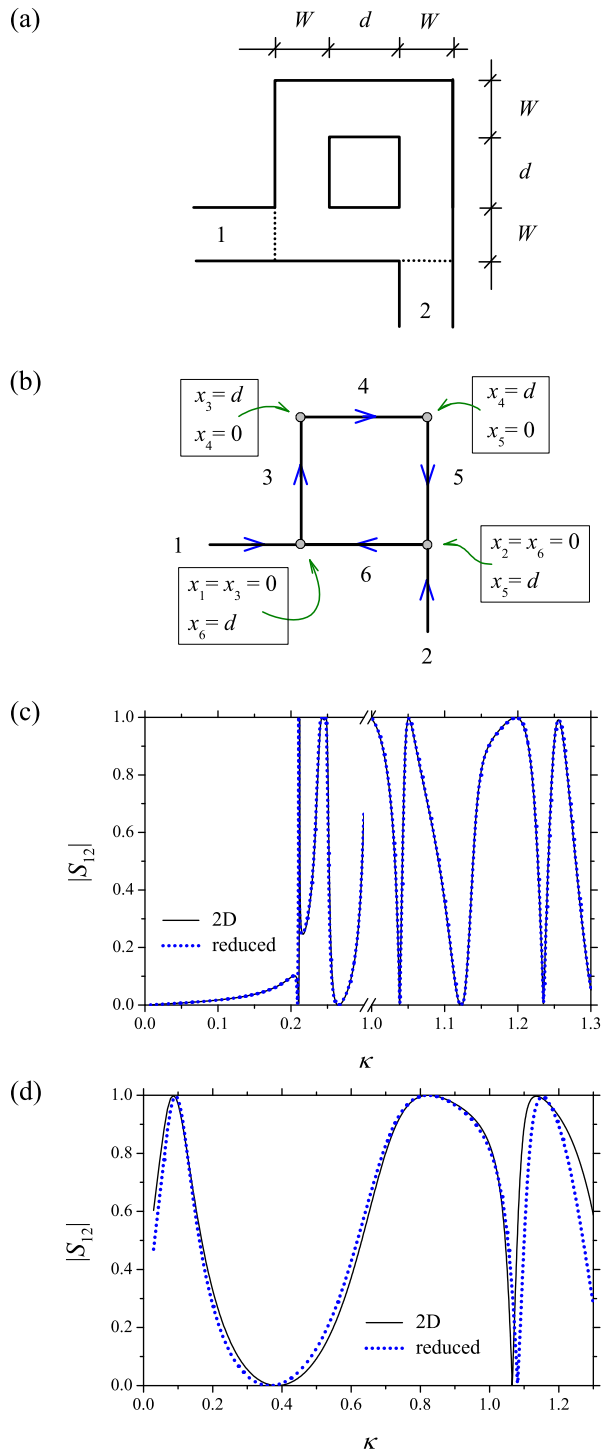}
\vspace {20cm}

\caption {See the caption in Fig.~\ref {loop1}.}

\label {loop2}
\end {figure}

The length of an exponential tail of the second transverse mode is of the
order of $|k^{(2)}|^{-1}$. The values of $|k^{(2)}W|^{-1}$ are 
approximately equal to 0.18 ($\kappa = 0$), 0.23 ($\kappa = 1.0$), 0.28
($\kappa = 1.3$), 0.37 ($\kappa = 1.5$), 0.48 ($\kappa = 1.6$), 0.96
($\kappa = 1.7$), and 1.56 ($\kappa = 1.72$), for the values of $\kappa$
given in the brackets \cite {Voo08a}.
It is seen that $|k^{(2)} W|^{-1}$ becomes large only when $\kappa$
approaches $\sqrt 3$. For $d=4W$, $|k^{(2)} d| \simeq 14 \gg 1$ at
$\kappa = 1.3$, and that justifies the reduction scheme in the entire
range of $\kappa$ in Fig.~\ref {loop1}(c) [and also Fig.~\ref {loop2}(c)
later in this section]. For $d=0.1W$, $|k^{(2)} d| \simeq 0.54$ at $\kappa
= 0$, and $|k^{(2)} d|$ is certainly not ``large'' for
Fig.~\ref {loop1}(d) [and also Fig.~ \ref {loop2}(d) later in this
section]. In spite of this, the scheme might still perform up to certain
precision, until it really starts to breakdown at $\kappa \sim 1.2$ as
seen in Fig.~\ref {loop1}(d) [see also Fig.~ \ref {loop2}(d)]. However,
its reliability in this regime of $\kappa$ is uncontrolled in general.

Another 2D wave guide as shown in Fig.~\ref {loop2}(a) is also
analyzed. This structure resembles the one in Fig.~\ref {loop1}(a),
except that one of the leads is rotated by 90-degree. The width of the
uniform sections are also denoted by $W$, and the distances between the
discontinuities are also denoted by $d$. A reduced structure for the guide
is shown in Fig.~\ref {loop2}(b), and the magnitude of the transmission
scattering amplitude $|S_{12}|$ is also plotted for both of the reduced
and unreduced structures in Figs.~\ref {loop2}(c) [for $d=4W$] and \ref
{loop2}(d) [for $d=0.1W$].

The connecting equations for the reduced structure in Fig.~\ref {loop2}(b)
are Eqs.~(\ref {node1}), (\ref {node2}), (\ref {node3}), and
\begin {eqnarray}
\displaystyle { { 1 + S _{\rm T} } \over {ik} }
\left[ \begin {array} {c}
\left. - \displaystyle {{d\psi_6} \over {dx_6}} \right| _{x_6=0}  \\
\left. ~ \displaystyle {{d\psi_2} \over {dx_2}} \right| _{x_2=0}  \\
\left. ~ \displaystyle {{d\psi_5} \over {dx_5}} \right| _{x_5=d}
\end {array} \right]
=
\left( 1 - S _{\rm T} \right)
\left[ \begin {array} {c}
\left. \psi_6 \right| _{x_6=0} \\ 
\left. \psi_2 \right| _{x_2=0} \\
\left. \psi_5 \right| _{x_5=d}
\end {array} \right]. 
\label {node4b}
\end {eqnarray}
The transmission scattering amplitude $S_{12}$ is found as in the
previous example. Likewise, it is seen in Fig.~\ref {loop2}(c) that the
reduction scheme is guaranteed to work in the $|k^{(2)} d| \gg 1$ regime.
Also, it is seen in Fig.~\ref {loop2}(d) that the scheme might still work
qualitatively or semi-quantitatively, when $\kappa$ is departed from this
regime.

Comparing the result in Fig.~\ref {loop1}(c) with that in Fig.~\ref
{loop2}(c), and the result in Fig.~\ref {loop1}(d) with that in Fig.~\ref
{loop2}(d), it is seen that scattering amplitudes can depend significantly
on the orientations of the branch guides at a discontinuity. This
indicates that reduction schemes with symmetric branch lines such as
those in Refs.~\cite {Kuh49,Gri53,RS53,VCT06} are not adequate for some
cases.

\begin {figure}

\includegraphics{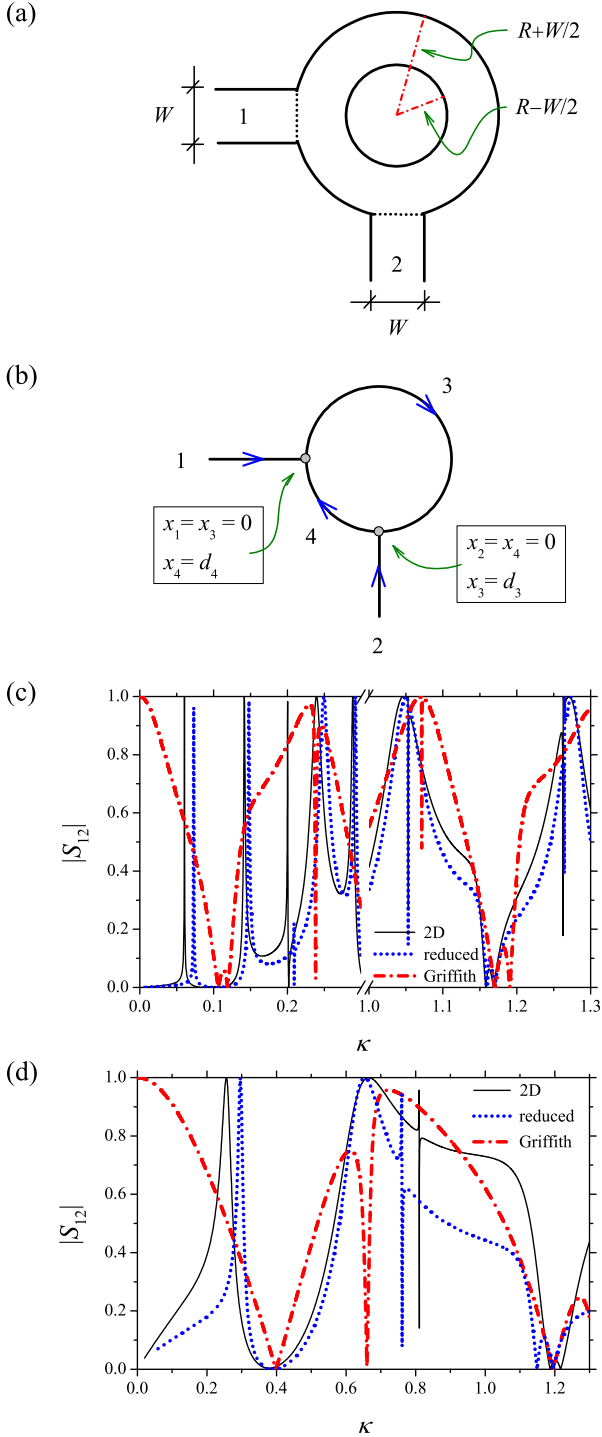}
\vspace {19cm}

\caption {
(Color online)
(a) A considered 2D annulus with two mutually perpendicular and radially
connected leads. 
The leads have the same width $W$. The inner radius $R-W/2$ and outer
radius $R+W/2$ of the annulus are defined as shown. 
(b) A reduced version of the annulus. A coordinate $x_\eta$ is
defined on line $\eta$ ($\eta = 1 - 4$) with a positive direction
indicated by an arrow on the line. The coordinates of the junctions
on the lines are shown in the boxes.
(c) For $R = 3W$, the magnitude of the transmission scattering amplitude
$|S_{12}|$ is plotted versus $\kappa$, for the original 2D annulus (solid
line) and the reduced system (dotted line for our scheme, and dash-dotted
line for the Griffith scheme). The $\kappa$ here is defined
by $\kappa \equiv k^{(1)} W / \pi$ (note that $k^{(1)}$ is defined in the
leads) or $\kappa \equiv kW/\pi$. 
(d) For $R = 0.8 W$, results for $|S_{12}|$ are also shown.
}

\label {ring}
\end {figure}

The third example is a 2D annulus structure with an inner and an outer
radii of $R - W/2$ and $R + W/2$ respectively, and two mutually
perpendicular leads of width $W$ radially connected to the annulus as
shown in Fig.~\ref {ring}(a). 
Figure \ref {ring}(b) shows a reduced version of it.
For the 2D annulus, we may follow a mode-matching full-wave treatment
formulated by Xia and Li \cite {XL02}.
In the annulus the wave function $\Psi_{\rm ann.}$ can be expanded by
radial and angular modes, $\Psi_{\rm ann.} (r,\theta) = \sum _{l=-M} ^M
\phi_l (Kr) e^{il\theta}$, where a radial mode is given by $\phi_l (Kr)
\equiv a_l J_l (Kr) + b_l Y_l (Kr)$, and $K = \sqrt {2 m_0 E}/\hbar$. The
$r$ and $\theta$ are the radial and angular coordinates respectively; and
the $J_l$ and $Y_l$ are the Bessel functions of the first and second kinds
respectively.
At the inner radius, $\phi_l | _{r=R-W/2} = 0$ for any $\theta$; At the
outer radius, $\Psi_{\rm ann.} | _{r=R+W/2} = 0$ when $\theta$ is away
from the leads, and $\Psi_{\rm ann.} | _{r=R+W/2} = \Psi_\eta$, when
$\theta$ is in the range of lead $\eta$.
In addition, the radial derivative $\partial \Psi_{\rm ann.} / \partial
r$ is equated with the longitudinal derivative $\partial \Psi_\eta / 
\partial x_\eta$ when $\Psi_{\rm ann.}$ and $\Psi_\eta$ meet at the outer
arc of the annulus. The difference between the straight transverse cuts of
the leads and the outer arcs of the annulus is neglected. The wave
functions in the leads and the annulus are hence matched, and one can get
a set of equations relating the coefficients of the modes in the different
regions. 
The transmission scattering amplitude $|S_{12}|$ for the 2D annulus is
plotted for $R = 3 W$ [Fig.~\ref {ring}(c)] and $R = 0.8 W$ [Fig.~\ref 
{ring}(d)].

%The results are obtained using 101 transverse modes in the leads ($N_\eta
%= 101$, where $\eta=1$ and 2) and 101 angular modes in the annulus ($M =
%50$). The differences between these results and those using $N_\eta =
%201$ and $M = 100$ are within the thicknesses of the data lines in the
%figures. 

To apply a reduced calculation to the reduced system in Fig.~\ref 
{ring}(b), note that for smoothly curved guides with small curvatures, the
back-scattering is small and the guides can be treated as reflectionless
for most purposes. Guides with constant curvatures are translational
invariant and indeed reflectionless, though the lengths and widths may not
be rigorously defined. The junctions are treated as T-junctions. For the
left junction,
\begin {eqnarray}
\displaystyle { { 1 + S _{\rm T} } \over {ik} }
\left[ \begin {array} {c}
\left. ~ \displaystyle {{d\psi_1} \over {dx_1}} \right| _{x_1=0}  \\
\left. - \displaystyle {{d\psi_3} \over {dx_3}} \right| _{x_3=0}  \\
\left. ~ \displaystyle {{d\psi_4} \over {dx_4}} \right| _{x_4=d_4}
\end {array} \right]
=
\left( 1 - S _{\rm T} \right)
\left[ \begin {array} {c}
\left. \psi_1 \right| _{x_1=0} \\ \left. \psi_3 \right| _{x_3=0} \\ \left.
\psi_4 \right| _{x_4=d_4}
\end {array} \right];
%\label {}
\end {eqnarray}
for the right junction,
\begin {eqnarray}
\displaystyle { { 1 + S _{\rm T} } \over {ik} }
\left[ \begin {array} {c}
\left. ~ \displaystyle {{d\psi_2} \over {dx_2}} \right| _{x_2=0}  \\
\left. ~ \displaystyle {{d\psi_3} \over {dx_3}} \right| _{x_3=d_3}  \\
\left. - \displaystyle {{d\psi_4} \over {dx_4}} \right| _{x_4=0}
\end {array} \right]
=
\left( 1 - S _{\rm T} \right)
\left[ \begin {array} {c}
\left. \psi_2 \right| _{x_2=0} \\ \left. \psi_3 \right| _{x_3=d_3} \\
\left. \psi_4 \right| _{x_4=0}
\end {array} \right].
%\label {}
\end {eqnarray}
We let $d_\eta = R \theta_\eta - W$, where $\theta_3 = 3 \pi/2$ and
$\theta_4 = \pi/2$, and the subtraction is to approximately exclude the
regions in the junctions [see region D in Fig.~\ref {discontinuities}(b)]. 
The radius $R = 3 W$ gives $d_3 \simeq 13.1 W$ and $d_4 \simeq 3.7 W$, and
$R = 0.8 W$ gives $d_3 \simeq 2.77 W$ and $d_4 \simeq 0.26 W$. Results for
$|S_{12}|$ are plotted in Figs.~\ref {ring}(c) and \ref {ring}(d).

If one analyzes the reduced system using the Griffith scheme \cite
{Kuh49,Gri53,RS53,Xia92} for a junction with three branches, 
the left junction has
\begin {eqnarray}
& \psi_1 |_{x_1=0} = \psi_3 |_{x_3=0} = \psi_4 |_{x_4=d_4} & {\rm and}
\\
& \left. {{d \psi_1} \over {dx_1}} \right|_{x_1=0} - 
\left. {{d \psi_3} \over {dx_3}} \right|_{x_3=0} + 
\left. {{d \psi_4} \over {dx_4}} \right|_{x_4=d_4} = 0,&
\end {eqnarray}
and the right junction has
\begin {eqnarray}
& \psi_2 |_{x_2=0} = \psi_3 |_{x_3=d_3} = \psi_4 |_{x_4=0} & {\rm and} \\
& \left. {{d \psi_2} \over {dx_2}} \right|_{x_2=0} + 
\left. {{d \psi_3} \over {dx_3}} \right|_{x_3=d_3} - 
\left. {{d \psi_4} \over {dx_4}} \right|_{x_4=0} = 0.&
\end {eqnarray}
Note that the equations at a junction are symmetric with respect to an
interchange of any two branches. Using the same $d_3$ and $d_4$,
results for $|S_{12}|$ are also plotted in Figs.~\ref {ring}(c)
and \ref {ring}(d). Comparing the results from our scheme and the
Griffith scheme with the result from the 2D calculation, it is seen that
the Griffith result is qualitatively different from the 2D result in
general, while our scheme captures the essential features in the 2D
result, especially in the case of longer guide sections [Fig.~\ref
{ring}(c)].

The lengths of the guide sections in Fig.~\ref {ring} are about the sizes
of those in Figs.~\ref {loop1} and \ref {loop2}, but the disagreement
between the 2D and reduced calculation results are seen to be more severe 
in Fig.~\ref {ring}, especially in the case of short guide sections
[Fig.~\ref {ring}(d)]. This is mainly due to vaguer notions of the
lengths and widths of the guide sections between the discontinuities, and
a stronger distortion of the shapes of the junctions from a ``T''
[Fig.~\ref {discontinuities}(b)], when the guide sections are curved and
short, in addition to a violation of the criterion $|k^{(2)}d| \gg 1$.
Comparing Figs.~\ref {ring}(c) and \ref {ring}(d), it is also seen that
our scheme performs better when the discontinuities are more apart.

\section {Concluding Remarks}
\label {conc}

This study has shown that the higher transverse modes in the
uniform-cross-sectional sections in wave guides can give only a minor 
effect on the low-energy properties of the guides. A scheme to remove
these modes and a criterion for the energy range ($|k^{(2)} d| \gg 1$) in
which the scheme applies have been proposed. In the scheme,
a reduced system and its corresponding system in a realistic space have a
sound relationship, and therefore the scheme may find more practical use
in the analyses of realistic wave guides than previously reported schemes
\cite {Kuh49,VCT06,Sha83}.

The electronic spin degrees of freedom may also be included into the
scheme by expanding the current one-mode $S$-matrices to two-mode
$S$-matrices, for the up and down spin channels, and in the same way as in
Sec.~\ref {examples}, the two-mode $S$-matrices are calibrated by 
full-wave calculations for systems in realistic spaces. 
Likewise, the precision of the scheme at short guide sections can also be
refined by including more transverse modes and using multi-mode 
$S$-matrices.

\newpage

%-------------------------------------------------------------------
%---------------------------------------------------------------

%---------------------------------------------------------------
%----------------------------------------------------------------

%-----------------------------------------------------------------
% . plot line width  =  1.5 - 4.
% . font size for insets  =  28
% . font size for numbers for axes  =  36.
% . font size for labels for axes  =  24
% . all SL/ST and Loop1/2 are from CST.
% . d=4W uses tetrahedral, and d=0.1W uses hexahedral meshing.
% (tetrahedral for a large freq range gave kinky arguments at low
% frequencies)
%---------------------------------------------------------------------

%---------------------------------------------------------------------
% see the remarks in fig. {loop1}
%---------------------------------------------------------------

%---------------------------------------------------------------------

%---------------------------------------------------------------

\end {document}